\documentclass[12pt]{article}
\usepackage{graphicx}

\usepackage{cite}

\begin{document}

\title{Physicists, Non Physical Topics, \\ and \\  Interdisciplinarity}

\author{ Serge Galam\thanks{serge.galam@sciencespo.fr} \\
CEVIPOF - Centre for Political Research, Sciences Po and CNRS,\\
1, Place Saint Thomas d'Aquin, Paris 75007, France}

\date{}
\maketitle

\begin{abstract}

Defining interdisciplinary physics today requires first a reformulation of what is physics today, which in turn calls for clarifying what makes a physicist. This assessment results from my forty year  journey arguing and fighting to build sociophysics. My view on interdisciplinary physics has thus evolved jumping repeatedly to opposite directions before settling down to the following claim: today physics is what is done by physicists who handle a problem the "physicist's way". However the training of physicists should stay restricted to inert matter. Yet  adding a focus on the universality of the physicist approach as a generic path to investigate a topic. Consequently, interdisciplinary physics should become a cabinet of curiosities including an incubator. The cabinet of curiosities would welcome all one shots papers related to any kind of object provided it is co-authored at least by one physicist. Otherwise the paper should uses explicitly technics from physics. In case a topic gets many papers, it would be moved to the incubator to foster the potential emergence of a new appropriate subfield of physics. A process illustrated by the subsection social physics in Frontiers in physics. 

\end{abstract}

Key words: Interdisciplinary physics, Sociophysics, history, epistemology

\section{Foreword}

During the last two decades physics has been increasingly applied to a large spectrum of topics outside the realm of inert matter. Or more precisely, physicists have been dealing with a large spectrum of topics outside the realm of inert matter substituting all kind of objects in place of the atom. Almost all  fields of knowledge are being visited by physicists applying their models and tools here and there  \cite{brazil1}. The trend gave rise to new subfields of physics like sociophysics and econophysics  \cite{ book-trend, book-stauffer, book-frank, review-fortunato, book}. 

While this on going jumping out of physics to tackle all sort of topics has definitively blown up the concept of interdisciplinary physics, no attempt has been made so far to design a clear frame to guide its current reshaping. This special issue addresses this challenge, to which I am contributing through what I have learned during my forty-year journey in arguing and fighting to build sociophysics among physicists, physics institutions, a few social scientists and one social science institution.

The case of sociophysics is emblematic since its maturation as an established new field of physics  \cite{frank} has been pondered by a series of successive contradictory steps, which can shed a constructive light about what should be interdisciplinary physics today as well as physics. 

During my difficult search for the appropriate frame to develop sociophysics my perspective on interdisciplinary physics went through several u-turns before I realized that indeed, sociophysics is the business of physicists and should be dealt within physics institutions as a new subfield of physics alongside traditional subfields as condensed matter or astrophysics. Some journals such as Frontiers in physics have taken this reality into account with the creation of a section named social physics.

\section{My assessment about physics and physicists}

Accordingly, before asking what interdisciplinary physics should be today, we need to answer first to the question of what is physics today? Broadening the case of sociophysics, my assessment is that today physics is what is done by physicists. Nevertheless, such a radical statement immediately raises the operative need to specify what makes up a physicist? 

My answer is that a physicist should still be trained the ``old way", i. e., studying the laws and properties of inert matter. The only addenda to the education of a physicist should thus be to enlighten what is related intrinsically to inert mater and what is related to the know how of physicists in tackling successfully the discovery of the laws governing inert matter. The focus being on the universality of the physicist approach as a generic path to investigate a topic. In particular, putting light on the scheme of the physicist modeling of a problem going first from crude approximations to step by step adding more features to get closer to the reality of the object being investigated. What I call the physicist's way.

While I emphasize the need to keep the physicist education mainly as it has been done in the last decades with the study of inert matter, that constraint does not forbid, on the contrary, enriching this education with the recent development of physics related for instance to complex systems out of equilibrium and nonlinear systems. However, I would favor that the extension of the subjects taught should be done only during the Master degree to preserve the "hard line" character of the physicist earlier stages of education.

But elaborating further with more details on the physicist education is beyond the scope of the present paper. The issue should be discussed collegially with simultaneous experimental courses to check and validate the various options.

\section{My proposal for interdisciplinary physics}

Once physics has been recast as what is done by physicists, I can suggest a new perimeter for the content and the role of interdisciplinary physics. interdisciplinary physics should become a cabinet of curiosities including an incubator. The cabinet of curiosities would collect and store unconventional topic papers produced by physicists with all one shot papers. And when a significant numbers of papers devoted to the same topic have accumulated, the topic would move to the incubator which will operate to strengthen it. Some, but not, of these topics will eventually become new subfields of physics as it happened with sociophysics \cite{brazil1, frank} and econophysics  \cite{book-stanley}.

Clearly, at this stage my proposal is very basic and would deserve more hints on how to consider its practical implementation. It mainly raises the following questions, which have been quoted by the referee:

\begin{itemize}
\item What measures of peer review benchmarks should be applicable at each stage of the reviewing process? 

\item Should there be dedicated journals devoted to the ``curiosities" papers or to the ``incubator" ones? 

\item How to classify them? 

\item By necessity, they would have low Impact Factors, so how would publishing in these dedicated courses journals influence author's careers? 

\item  Should there be special grant schemes, especially in the incubator phase?

\end{itemize}

These questions by the referee are very interesting and essential to put together my proposal but they are far beyond the goal of the present paper. I thing those practical measures, especially making suggestions for promoting related publications, education, and dissemination should also be discussed collegially and in another frame.

\section{The early times: envisioning a physicist invasion}

I have reported a personal testimony on the earlier stage of sociophysics, which was initiated in the late seventies and beginning eighties \cite{testimony}. It is worth to recall that the historical context of physics at that time was a kind of golden age for condensed matter and statistical physics. In 1971, 

The concepts and tools of the renormalization group techniques opened a wide area of intense and active research in physics with hundreds of PhD done on the topic.

However, at the same time, most of tenured position were already held by young physicists with no retirement in sight and only a very little number of openings were available for the multitude of new trained doctors. Being myself in that last category I foresaw for the near future the formation of a community of frustrated young and skilled physicists. I wrote a short note entitled "Physicists' misery"  \cite{mise}. I had the idea to define this class of out of job young and frustrated physicists as a revolutionary catalyst to embrace  all the various fronts of research out of physics  \cite{frust1, frust2, revo}. But at the same time being aware of colonialism dynamics I also set a dialectical frame which envisions in a second step, a fight back from the ``colonized" fields against the imperialism of physics, which in turn would push these physicists back to their original field, physics, enriched from their mixing with those ``foreign" fields \cite{imp}.

Unfortunately,  my view stalled totally staying utopian before it eventually became a reality at the end of twentieth century. The burst of econophysics and sociophysics subscribes to this global picture elaborated in a series of 10 papers, half of them co-authored with Pierre Pfeuty  \cite{mise, frust1, frust2, revo, imp, philo1, philo2, philo3, philo4, philo5}.

\section{Publish outside of physics in social science mathematical journals}

During about two decades I had to fight against local physicists determining if sociophysics could  be published with an affiliation to the local physics department where I was.  In the last ten years or so the fight has been closed with now acceptance of sociophysics works in most physics journals and institutions.
 
It is impossible for young and mid age researchers to even imagine the harsh hostility deployed by physicists against any deviation from the study of inert matter during the late seventies till the nineties. Indeed, it is remarkable to note that this not very glorious part of the beginning of sociophysics is being obliterated with current revisionist rewriting of the history of sociophysics tracing its roots back to a few centuries ago to some social science authors  \cite{wiki-socio, stauffer-history, ball}

The fight against this ``heretic" deviation from physics to study human behavior has culminated at the end of the seventies at Tel-Aviv University. There, at the physics department with Yuval Gefen and Yonathan Shapir we wrote a founding paper for sociophysics quoting for the first time the name sociophysics in the title. Amazingly, our typed manuscript has been then confiscated by the chairman of the physics department. The move was supported by most physicists from the department who were seeing our manuscript as a threat which could jeopardize the international reputation of the department. Thanks to Alexander Voronel support, a Jewish physicist newly arrived from Soviet Union, we got back our manuscript \cite{testimony}.

Happy to have recovered our work, the three of us had then to face the difficult choice of which journal to submit our manuscript. It even did not cross our mind to consider a physics journal. It quite obvious for the three of us to look for a social science journal with mathematical background. We ended up sending our manuscript to the Journal of  Mathematical Sociology where it was latter accepted after many back and forth with several round of referees  \cite{strike}. 

Once in New York I committed another sociophysics work, which I then submitted to the Journal of Mathematical psychology. Again, I went trough a series of exchanges with the referee before the paper was accepted and published  \cite{voting1}. 

I can imagine how awkward it has been for the various referees to read and comment those unidentified papers by physicists. But yet, to their credit they did a positive job with constructive comments with the papers being eventually published.

\section{A one shot publication in a physics journal}

Incidentally, at  the end of the eighties I had a chance to meet with Dietrich Stauffer at Tel-Aviv University where we were both visitors. We happened to share an office and Stauffer got very interested in my sociophysics approach and my paper on bottom up democratic structures \cite{voting1}. He then suggested me to submit a follow-up paper to the Journal of Statistical Physics where he was an Editor. And I did with a paper eventually published in 1990  \cite{voting2}. 

This publication is significant since if I am not mistaken, it has been the first physics journal ever, to publish a sociophysics paper, thanks to the handling of the paper by Stauffer. It is noticeable that the paper stays a one shot event for many years. n addition, It is of importance to underline that once the paper has been accepted for publication, I got a handwritten letter (no email in these days) from the Editor in Chief Joel Lebowitz, who pointed to me his strong disagreement with the approach of the paper. To his credit, he wanted to express his personal negative view about the publication but did not oppose it  since the paper had gone through the refereeing process successfully. 

This one shot intrusion in a physics journal stayed unique for many years and amazingly, twenty two years latter, in 2013 the Journal of Statistical Physics has devoted a special issue to mathematical social modeling  \cite{voting3}.

\section{A strategy change: setting a collaboration with a social scientist}
 
Having got almost zero feedback from my papers published in mathematical social science journals and not more from my publication in the Journal of Statistical Physics I figured that I should find a social scientist who could get involved in a collaboration to launch a research along sociophysics. That could be the path to reach a community of social scientists.

That goal was achieved collaborating with Serge Moscovici, a French psycho-sociologist, with who we elaborate a theory of group decisions making  with a series of co-authored papers published in the European Journal of Social Psychology  \cite{mosco1, mosco2, mosco3}. The papers did generate some feedback but not at a significant level. In addition, the collaboration stayed restricted to one person with no much of opportunity to get beyond. I thus decided to stop our collaboration and go back to research on my own.

\section{Focus on physics journals}

At the same time, I met again with Stauffer who reiterated his advice to submit papers in physics journals, in particular in Physica A where he was also an Editor. Indeed, thanks to him and Gene Stanley who was the chief Editor, Physica A became in the 2000s the main journal for publications in sociophysics and econophysics. I thus  published  there several papers and also in other physics journals. With time and more or less difficulties all physics journals eventually opened their pages to sociophysics papers. That opening took about twenty years or so to be completed.

\section{Join a social science lab within a scientific institution}
 
Yet, I felt that something could be improved to foster sociophysics. So, when later on I was offered to join the Research Center for applied Epistemology (CREA) at Ecole Polytechnique, I thought that will be the perfect place to be in tune with a full lab.  

This center of social science had the double peculiarity of being hosted by a scientific institution, the Ecole Polytechnique,  and composed mostly with researchers having a strong mathematical background. I eventually joined the CREA thinking it would also be appropriate to build some classes for the engineering students' training program at the School. My preconception was that engineering students would be immediately attracted to physical modeling of the social reality. 

I was wrong not due to the lack of interest from the students but from the management's attitude with had no interest at all to modify their teaching programs along that direction. 

\section{Join a political science lab in a social science institution}

After nine year at CREA  I made another bifurcation joining  the CEVIPOF (Centre for Political Research) at Sciences Po, a social science university in Paris. This time no mathematical background for most of the political scientists, indeed I became there the first physicist ever.

My new preconception was that it would be more ``natural" to rise interest among researchers by adding a new approach to their current topics of interest than before, when I thought adding a new topic to be investigated by the current tools which were being used.

My adventure at CEVIPOF and Sciences Po has been quite successful despite both the huge epistemological challenges and the cultural differences. I got well accepted in the group and was also offered to give a class to Master students to introduce my sociophysics modeling. 

\section{Staying with physicists in physics institutions}

My experience at CEVIPOF has been and is quite stimulating with a good deal of discussions and exchanges with my colleagues. I add my sociophysics contribution to the topics of the lab, in particular the study of opinion dynamics in connection to predicting electoral events  \cite{brazil2, trump2}. However, I kept publishing and collaborating mostly with physicists \cite{taksu, radi, andre-serge}. 

At the moment I am an Emeritus Senior researcher and I am happy to keep on working at CEVIPOF. But at the same time, I don't see my case as being replicated and extended by bringing in more physicists.

\section{Conclusion}

Before concluding, I would like to stress the fact I did not intend to address here the fundamental question on how to to bridge the gap between social sciences and physics in order to make sociophysics an effective predictive tool to forecast some political events like election outcomes. I am aware that an atom and a human being have not much in common. However, the aim of our physicist approach is to succeed in extracting some salient features of human beings, which are driving some specific social dynamics or trends. We do not intend to neither substitute to social science nor to become social scientists. Our aim is to build another field to explore human behavior in addition to the paradigms of social sciences. We are facing both limitations and challenges and that is what makes the scientific adventure exciting with still a long way to go and search. My statement applies possibly to other new emerging fields from physicists.
 
 To keep on exploring this novel path of research, rich of my forty year journey across a series of academic institutions in several different countries, I would advocate``No invasion, of other fields, no getting out of physics environnments, stay in physics laboratories" and there enlarge the topics to be dealt by physicists. Indeed, that is an acknowledgement of reality  \cite{sen, nuno, malarz, andre, nguyen, red, kasia1, cheon, celia, bagnoli, zanette, belief, iglesias, mobilia, fasano, mariage, tessone, deffuant}. However, such an explication is not without consequences because it will allow to have classes of sociophysics in the curriculum of physics studies and it will also allow to have PhD in sociophysics. It also worth to notice that the modeling of social systems is attracting a good deal of mathematicians \cite{orso1, orso2, bello, lanch}.

Having recast physics as being what is done by physicists, interdisciplinary physics would be reshaped to become the outdoor of physics. It would welcome all no subfield identified works with both a cabinet of curiosities and a new field incubator. Acceptance to label a paper as Interdisciplinary physics should be subject to have at least one co-author physicist provided this physicist has being acting the physicist's way. Otherwise the requirement would be to have non physicists who have used clearly identified tools and technics from physics. 

One shot papers would be part of a cabinet of curiosities collecting unconventional topic papers produced by physicists. In addition, if many papers start to accumulate with the same unconventional topic, the topic would go to a subpart defined as an incubator for new topics. The incubator would operate to foster a possible new emerging topic. Some of those topics but not all of them will eventually become new subfields of physics as it happened with sociophysics and econophysics. In short, interdisciplinary physics could become a leading hub for innovative and multi object research within physics journals and institutions.

My paper is only one contribution to the crucial Issue addressed by this Special Issue. To better shape our future challenging frames of research, we need more exchanges, more discussions, more experimental classes, more forums. This Special Issue is a first step in that direction.

\section*{Acknowledgment}

I would like to thank the anonymous referee, who has raised several relevant questions with constructive comments. 



\end{document}